\documentclass[english,prl, two column]{revtex4-1}
\usepackage{lmodern}

\usepackage[T1]{fontenc}
\usepackage[latin9]{inputenc}
\usepackage{color}
\usepackage{babel}
\usepackage{amsmath}
\usepackage{amssymb}
\usepackage{graphicx}
\usepackage{subscript}
\usepackage[unicode=true,pdfusetitle,
 bookmarks=true,bookmarksnumbered=false,bookmarksopen=false,
 breaklinks=false,pdfborder={0 0 0},pdfborderstyle={},backref=false,colorlinks=true]
 {hyperref}
\hypersetup{
 colorlinks,linkcolor=red,citecolor=blue}
\usepackage{breakurl}
\begin{document}

\title{Superconducting cavity electro-optics: a platform for coherent photon
conversion between superconducting and photonic circuits}

\author{Linran Fan, Chang-Ling Zou, Risheng Cheng, Xiang Guo, Xu Han, Zheng
Gong, Sihao Wang}

\affiliation{Department of Electrical Engineering, Yale University, New Haven,
Connecticut 06511, USA}

\author{Hong X. Tang}
\email{hong.tang@yale.edu}

\selectlanguage{english}%

\affiliation{Department of Electrical Engineering, Yale University, New Haven,
Connecticut 06511, USA}

\maketitle
\textbf{Leveraging the quantum information processing ability of superconducting
circuits and long-distance distribution ability of optical photons
promises the realization of complex and large-scale quantum networks.
In such a scheme, a coherent and efficient quantum transducer between
superconducting and photonic circuits is critical. However, such quantum
transducer is still challenging since the use of intermediate excitations
in current schemes introduces extra noise and limits bandwidth. Here
we realize direct and coherent transduction between superconducting
and photonic circuits based on triple-resonance electro-optics principle,
with integrated devices incorporating both superconducting and optical
cavities on the same chip. Electromagnetically induced transparency
is observed, indicating the coherent interaction between microwave
and optical photons. Internal conversion efficiency of $25.9\pm0.3\%$
has been achieved, with $2.05\pm0.04\%$ total efficiency. Superconducting
cavity electro-optics offers broad transduction bandwidth and high
scalability, and represents a significant step towards the integrated
hybrid quantum circuits and distributed quantum computation.}

\vbox{}

\noindent\textbf{\large{}Introduction}{\large \par}

The hybrid approach of combining superconducting and photonic quantum
technologies promises to realize large scale quantum networks \citep{schoelkopf2008wiring,o2009photonic,lvovsky2009optical,xiang2013hybrid}.
In superconducting quantum circuits, the low-loss single quanta nonlinearity
at microwave frequencies inherent to Josephson effect allows efficient
and fast quantum operations \citep{clarke2008superconducting}. However,
it is challenging to directly transmit quantum states at microwave
frequencies over long distance due to the high attenuation and thermal
noise at room temperature. On the other hand, optical photons show
complementary features. The weak single photon nonlinearity prevents
the development of high fidelity quantum gates at optical frequency
\citep{kok2007linear}. However, low decoherence and dissipation rates
make optical photons the ideal information carrier for quantum communication
\citep{o2009photonic,lvovsky2009optical}. As a result, it is beneficial
to develop hybrid quantum platform where quantum information is processed
by superconducting circuits, and transmitted with optical photons.
Thus, the quantum transducer which can coherently interface superconducting
and photonic circuits with high conversion efficiency is highly demanded
\textbf{\citep{andrews2014bidirectional,bochmann2013nanomechanical,balram2016coherent,barzanjeh2012reversible,lecocq2016mechanically,tsang2010cavity,tsang2011cavity,Huang2015,javerzac2016chip,Rueda:16,soltani2017efficient,hafezi2012atomic,Hisatomi2016,williamson2014magneto}}.

Coherent conversion of photons between microwave and optical frequencies
has been proposed utilizing various intermediate excitations, including
collective spin in atom ensembles \citep{hafezi2012atomic}, phonon
in electro-optomechanics \citep{barzanjeh2012reversible,bochmann2013nanomechanical,andrews2014bidirectional,balram2016coherent},
and magnon in magneto-optics \citep{Hisatomi2016,williamson2014magneto}.
Currently, the highest conversion efficiency is demonstrated based
on electro-optomechanical systems where a compliant mechanical resonator
couples to microwave and optical cavities simultaneously \citep{andrews2014bidirectional}.
The use of intermediate low-frequency (MHz) excitations inevitably
introduces extra noise channels, limits conversion bandwidth, and
complicates operation with impedance matching between optical and
microwave ports. The electro-optic approach can overcome these obstacles
by excluding intermediate excitations in the conversion process, as
proposed by Tsang recently \citep{tsang2010cavity,tsang2011cavity}.
Great improvement of electro-optic coupling strength is proposed by
Javerzac-Galy et al. utilizing coplanar microwave structure and integrated
optical resonators, showing the feasibility of near-unity conversion
efficiency with practical device parameters \citep{javerzac2016chip}.
Even though the conversion efficiency of electro-optical systems has
been improved dramatically, it is still limited to \textasciitilde{}0.1\%
due to the large ohmic loss of non-superconducting material, and small
coupling rates resulting from large mode volumes \citep{Rueda:16}.
Moreover, the coherence and bidirectionality of the conversion process
remain to be proved.

In this paper, we report the experimental demonstration of the coherent
conversion between microwave and optical photons based on the electro-optic
effect within a hybrid superconducting-photonic device, where planar
superconducting resonators are integrated with aluminum nitride (AlN)
optical cavities on the same chip. We observe the electromagnetically
induced transparency effect in electo-optic systems, as a signature
of coherent conversion between microwave and optical photons. Internal
conversion efficiency of \textbf{$25.9\pm0.3\%$} and on-chip efficiency
of \textbf{$2.05\pm0.04\%$} are realized. A major challenge we have
addressed is to realize the energy and phase conservation of the triple-resonance
condition with ultra-small mode volumes, boosting the pump photon
number and vacuum coupling rate simultaneously to enhance the coherent
conversion. Moreover, our device is ready to incorporate other superconducting
and photonic quantum devices on the same chip, providing the scalable
platform for hybrid quantum network.

\vbox{}

\noindent\textbf{\large{}Results}{\large \par}

\begin{figure}
\begin{centering}
\includegraphics{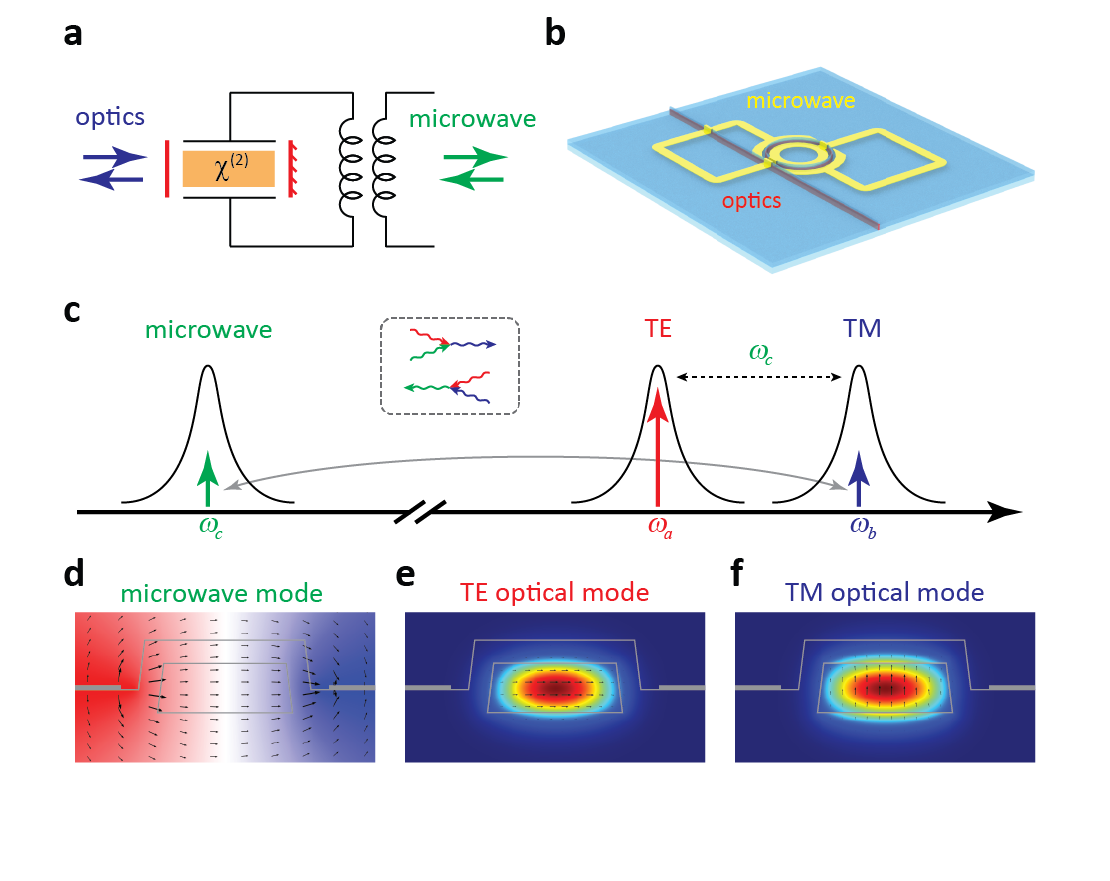}
\par\end{centering}
\caption{\textbf{Coherent conversion with cavity electro-optics. }(\textbf{A})\textbf{
}Schematic of cavity electro-optic systems. The optical cavity is
made of materials with Pockel nonlinearity ($\chi^{(2)}$), and placed
in the capacitor of the LC circuit. At the same time, optical and
microwave cavities are coupled to optical and microwave bus waveguides
respectively. (\textbf{B})\textbf{ }Integrated superconducting cavity
electro-optic device. The red part is the optical cavity and coupling
waveguide, and the yellow part is the superconducting microwave cavity.
A buffer layer (semi-transparent) is placed between optical devices
and the superconducting cavity to prevent metallic absorption of optical
photons. (\textbf{C})\textbf{ }Diagram of frequencies in the conversion
process. Strong control light is applied to the TE optical mode (pump
mode), and photons can be converted between the microwave mode and
TM optical mode (signal mode). Microwave photons are converted to
optical photons through sum frequency generation, and optical photons
are converted to microwave photons through difference frequency generation
as shown in insets. The mode distribution in the cross-section is
shown for the microwave mode (\textbf{D}), TE optical mode (\textbf{E}),
and TM optical mode (\textbf{F}). Arrow direction and length represent
the electric field direction and strength in log scale respectively.
Colors in (\textbf{D}) represent the voltage distribution, and colors
in (\textbf{E})\textbf{ }and\textbf{ }(\textbf{F})\textbf{ }represent
the energy density. In simulation, the optical waveguide is $2\,\mathrm{\mu m}$
wide and $800\,\mathrm{nm}$ thick, and the sidewall angle is $8\,$deg.
The distance between microwave electrodes is $2.8\,\mathrm{\mu m}$.
The material boundary is plotted in grey.}

\label{Fig1}
\end{figure}

The principle of quantum transducers based on superconducting cavity
electro-optics is illustrated in Fig.$\,$\ref{Fig1}A. The optical
cavity, consisting of materials with Pockel nonlinearity, is placed
inside the capacitor of the LC microwave resonator. The electric field
across the capacitor changes the refractive index of the optical cavity,
thus modulates the optical resonant frequency. Reversely, modulated
optical fields can generate microwave field due to the optical mixing
(rectification) in Pockels materials \citep{bass1962optical}. To
implement the quantum transducer, we use integrated optical microring
cavities made of AlN, which supports low loss optical modes and provides
high electro-optic coefficients simultaneously \citep{xiong2012aluminum}
(Fig.$\,$\ref{Fig1}B). Superconducting microwave resonators are
placed on top of a thin buffer layer, and the capacitor shape is designed
to match the optical cavity to maximize the field overlap between
microwave and optical modes \citep{javerzac2016chip,soltani2017efficient}.

\begin{figure*}
\begin{centering}
\includegraphics{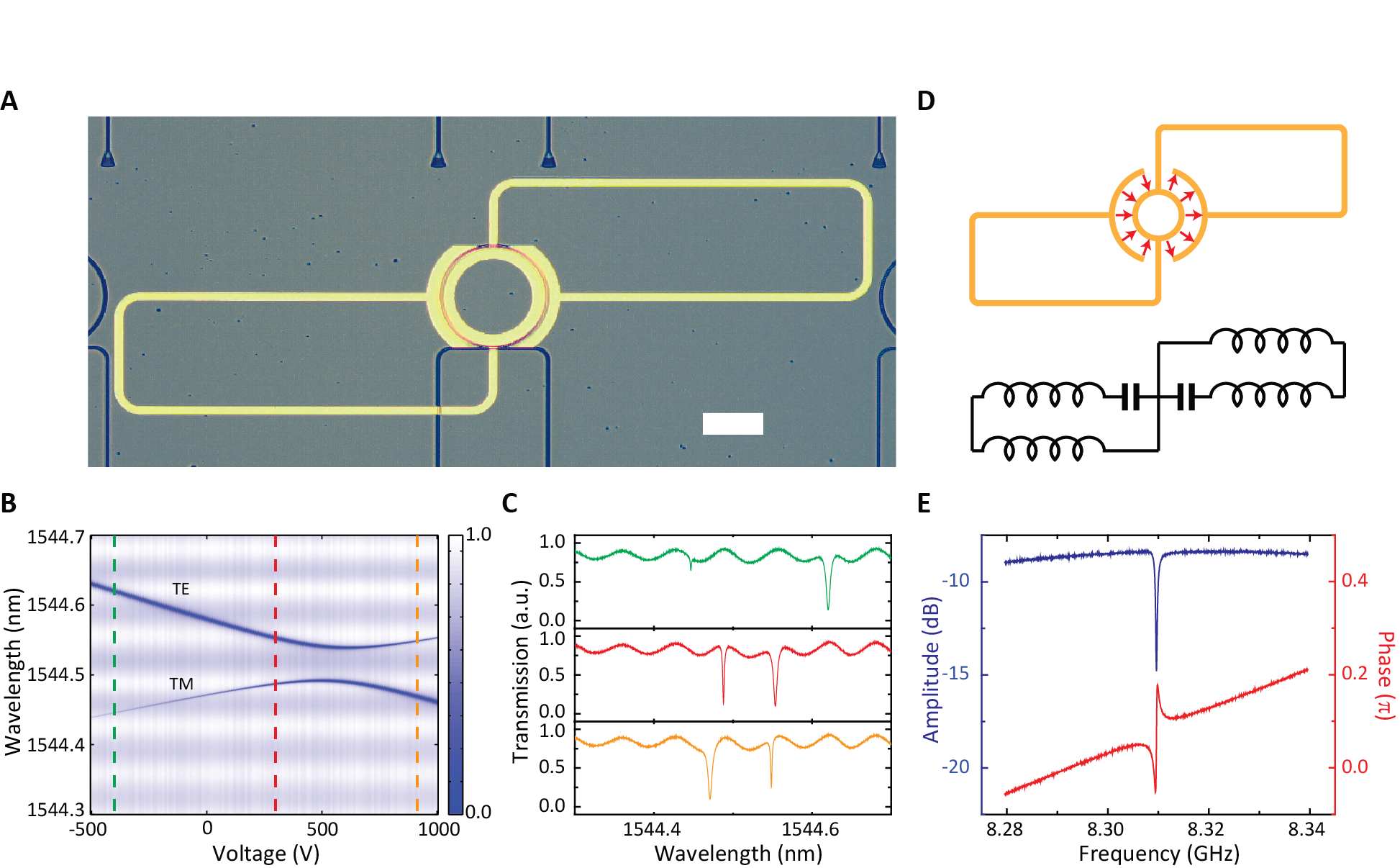}
\par\end{centering}
\caption{\textbf{Integrated superconducting cavity electro-optic device.} (\textbf{A})
Optical image of the superconducting cavity electro-optic device.
The scale bar is $100\,\mathrm{\mu m}$. (\textbf{B}) TE optical spectrum
with different DC voltages. The azimuthal number difference between
the TE and TM optical modes is 1. The mode anti-crossing gap is $2g_{\mathrm{x}}\sim6.1\,\mathrm{GHz}$,
and the original dissipation rates for TM and TE optical modes without
optical mode mixing are 190$\,$MHz and 480$\,$MHz, respectively.
(\textbf{C}) TE optical spectrum with DC voltages of -400$\,$V (green),
300$\,$V (red), 900$\,$V (orange), corresponding to the green, red,
and orange dash lines in (\textbf{B}) respectively. (\textbf{D}) Schematic
of the microwave resonator and electric field distribution of the
microwave mode, as well as the equivalent circuit. (\textbf{E})\textbf{
}Measured reflection spectrum of the microwave cavity.}

\label{Fig2}
\end{figure*}

We employ the triple-resonance scheme to enhance the coupling as shown
Fig.$\,$\ref{Fig1}C, with the interaction Hamiltonian written as
\[
H_{\mathrm{I}}=\hbar g_{\mathrm{eo}}(ab^{\dagger}c+a^{\dagger}bc^{\dagger})
\]
where $a$, $b$, $c$ are the annihilation operators for the optical
pump and signal modes, and microwave mode respectively, and $g_{\mathrm{eo}}$
is the vacuum electro-optic coupling rate. Cavity electro-optic systems
with triple resonances have been proposed and demonstrated recently
\citep{javerzac2016chip,Rueda:16}. However, the device geometry is
limited to above several millimeters, as the pump and signal optical
modes are from the same mode group, and the free spectral range (FSR)
needs to match the microwave frequency. Large mode volume inevitably
leads to small vacuum coupling rate, thus low conversion efficiency
(Supplementary Section I). In contrast, our integrated approach uses
the transverse-electric (TE) and transverse-magnetic (TM) optical
modes as pump and signal modes respectively (Fig.$\,$\ref{Fig1}E
\& F), whose frequency difference equals the microwave frequency \citep{ilchenko2002sub,ilchenko2003whispering,savchenkov2009tunable,strekalov2009microwave}.
Thus the device size and mode volume can be further reduced without
the limitation imposed by FSR. In this case, $r_{13}$ electro-optic
coefficient is used, which also enables the use of TE-polarized microwave
mode, thus the heterogeneous integration of planar microwave resonators
with optical cavities (Fig.$\,$\ref{Fig1}D). During experiments,
a strong coherent field is applied to the pump mode ($a$) to stimulate
the coherent coupling between signal mode ($b$) and microwave mode
($c$), and photons can be bidirectionally converted between optical
and microwave frequencies with on-chip efficiency 
\begin{equation}
\eta=\frac{\kappa_{b,\mathrm{ex}}}{\kappa_{b}}\frac{\kappa_{c,\mathrm{ex}}}{\kappa_{c}}\frac{4C}{(1+C)^{2}},
\end{equation}
where $\kappa_{b,\mathrm{ex}}$, $\kappa_{b}$, $\kappa_{c,\mathrm{ex}}$,
$\kappa_{c}$ are the external coupling and total loss rates for signal
and microwave modes respectively, and $C=\frac{4n_{a}g_{\mathrm{eo}}^{2}}{\kappa_{b}\kappa_{c}}$
is the cooperativity with $n_{a}$ the photon number in the pump mode
(Supplementary Section I).

\begin{figure*}
\begin{centering}
\includegraphics{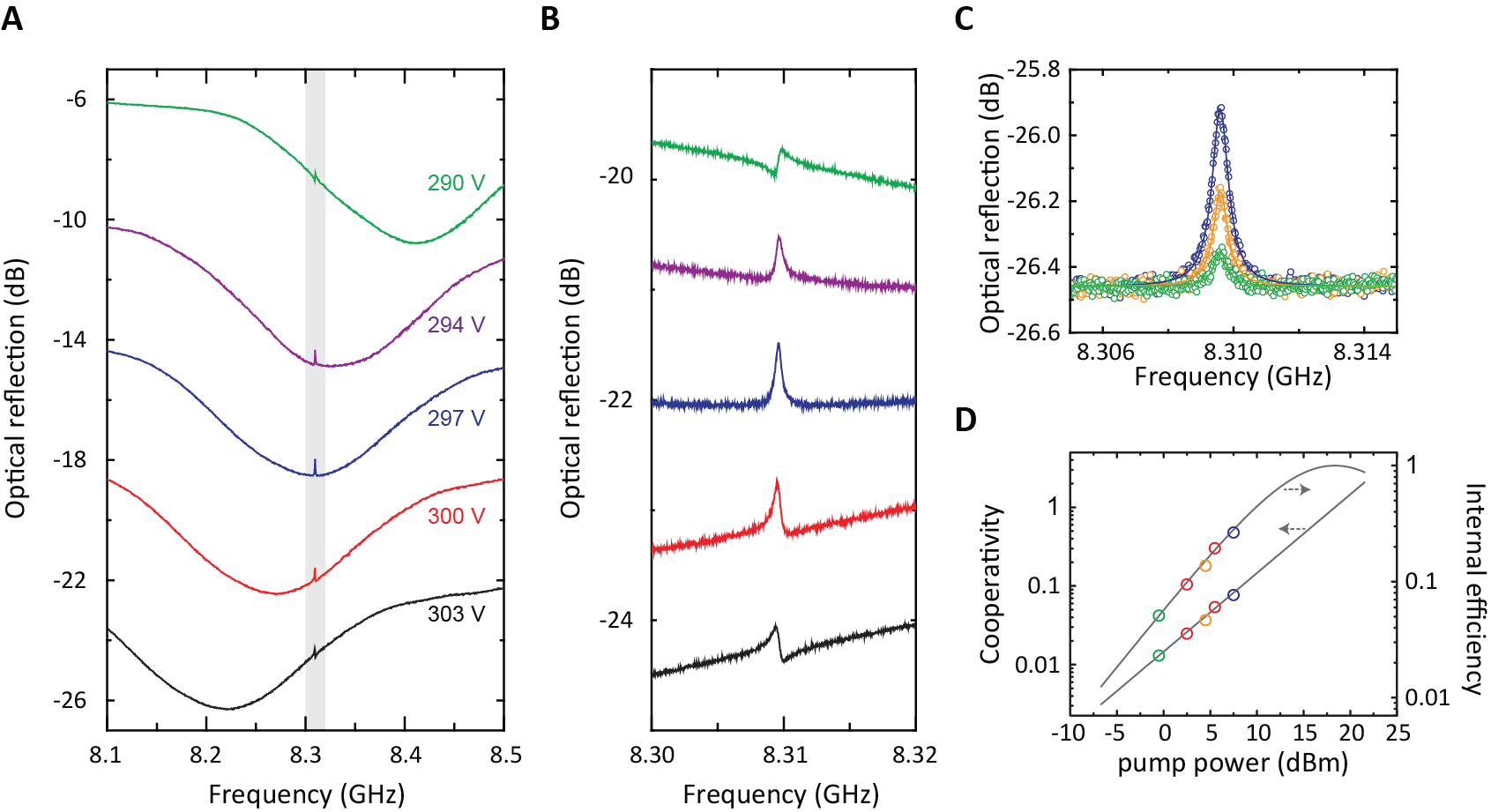}
\par\end{centering}
\caption{\textbf{Electromagnetically induced transparency with cavity electro-optics.}
(\textbf{A}) Measured optical reflection spectrum as a function of
the modulation frequency. (\textbf{B}) Zoom-in of the optical reflection
spectrum centered at the transparency window. Each spectrum in (\textbf{A})\textbf{
}and (\textbf{B}) corresponds to a different DC voltage, thus different
frequency detuning between pump and signal modes. Spectrums are offset
for clarity. (\textbf{C}) Transparency window with the control light
power of 8$\,$dBm (blue), 5$\,$dBm (orange), and 0$\,$dBm (green).
Circles are measured data, and solid lines are fitted spectra with
Eq.$\,$S13 in Supplementary Section I. (\textbf{D}) Cooperativity
and internal conversion efficiency versus control light power. The
blue, orange, and green points correspond to the blue, orange, and
green curves in (\textbf{C}) respectively. Grey lines are the fitted
result based on measured data.}
\label{Fig3}
\end{figure*}

In experiments, the optical cavity is fabricated from an AlN layer
on a silicon dioxide cladding on silicon wafer (Fig.$\,$\ref{Fig2}A).
The optical ring cavity has a radius of $120\,\mathrm{\mu m}$, and
cross-section of $2.0\,\mathrm{\mu m}\times0.8\,\mathrm{\mu m}$ (See
Supplementary Section II for fabrication procedure and device cross-section).
An azimuthal number difference of 1 between the pump and signal optical
modes is chosen to mitigate the optical mode mixing induced by the
non-vertical waveguide sidewalls (See Supplementary Section III for
the identification of azimuthal number difference, and Supplementary
Section IV for the influence of TE/TM mode mixing). In Fig.$\,$\ref{Fig2}B,
we present the transmission of TE input light, where mode anti-crossing
is observed arising from the structural asymmetry and fabrication
imperfection. As TE and TM optical modes have opposite electro-optic
coefficients, DC voltage is applied across the device to tune the
frequency difference precisely in a wide spectrum range. Compared
with using optical modes from the same group \citep{javerzac2016chip,Rueda:16},
the large frequency tuning range of our device makes it easy to accommodate
different microwave frequencies within a single device.

The microwave resonator is made of NbTiN superconducting film with
critical temperature around $14\,\mathrm{K}$. The device is placed
in a cryostat and cooled down to 2$\,$K, and the device surface is
covered by superfluid helium to introduce fast heat dissipation \citep{sun2013nonlinear}.
To allow electro-optic phase matching, it is important to shape the
microwave resonator to have azimuthal number of 1, to match the azimuthal
number difference between the pump and signal modes (Fig.$\,$\ref{Fig2}D
\& Supplementary Section V). The capacitance part of the microwave
resonator has a radius of $120\,\mathrm{\mu m}$ to match the optical
cavity, and the distance between electrodes is $2.8\,\mathrm{\mu m}$.
Each inductance arm has a length of $1.5\,\mathrm{mm}$, allowing
far-field magnetic coupling to an off-chip loop probe for broadband
microwave signal input and readout. The microwave mode has resonance
around $\omega_{c}/2\pi=8.31\,\mathrm{GHz}$ with the decay rate of
$\kappa_{c}/2\pi=0.55\,\mathrm{MHz}$ at $2\,\mathrm{K}$ (Fig.$\,$\ref{Fig2}E).
When the DC voltage is tuned to $297\,\mathrm{V}$, the frequency
difference between pump and signal modes is also around $8.31\,\mathrm{GHz}$
(Fig.$\,$\ref{Fig2}C). Therefore, the phase matching and energy
conservation are fulfilled simultaneously. 

The coherent conversion of our device is first characterized with
optical reflection spectrum. Strong control light is applied to the
pump mode, and a weak probe light, derived from the control light
by single side-band modulation, is sent to the signal mode (Supplementary
Section VI). No obvious temperature change of the superconducting
microwave resonator is observed (Supplementary Section VII). Figure$\,$\ref{Fig3}A
presents the probe light transmission spectrum sweeping across the
signal mode, with a fixed control light on resonance with the pump
mode. By tuning the DC voltage, the broad Lorentzian dip corresponding
to the signal mode is shifted, with a sharp modification of the spectrum
at a fixed frequency $\omega=\omega_{c}$. This modification originates
from the destructive interference between two pathways for the probe
light: directly passing through the optical cavity, and converting
to microwave photons and then back to optical photons \citep{weis2010optomechanically,safavi2011electromagnetically}.
As shown in the enlarged spectrum (Fig.$\,$\ref{Fig3}B), when the
signal mode frequency matches $\omega_{c}$, there is a sharp transparency
window, with the bandwidth equal to $\left(1+C\right)\kappa_{c}=(0.59\pm0.01)\,\mathrm{MHz}$.
If the signal mode is detuned, the interference gives an asymmetric
Fano-shape spectrum. By fitting the transparency window, the conversion
cooperativity can be extracted, and internal conversion efficiency
can be inferred (Fig.$\,$\ref{Fig3}C \& D) \citep{hill2012coherent}.
An internal efficiency as high as $(25.9\pm0.3)\%$ is achieved with
our device under 8 dBm control light power (14 dBm total off-chip
optical power with 6 dB insertion loss).

\begin{figure}
\begin{centering}
\includegraphics{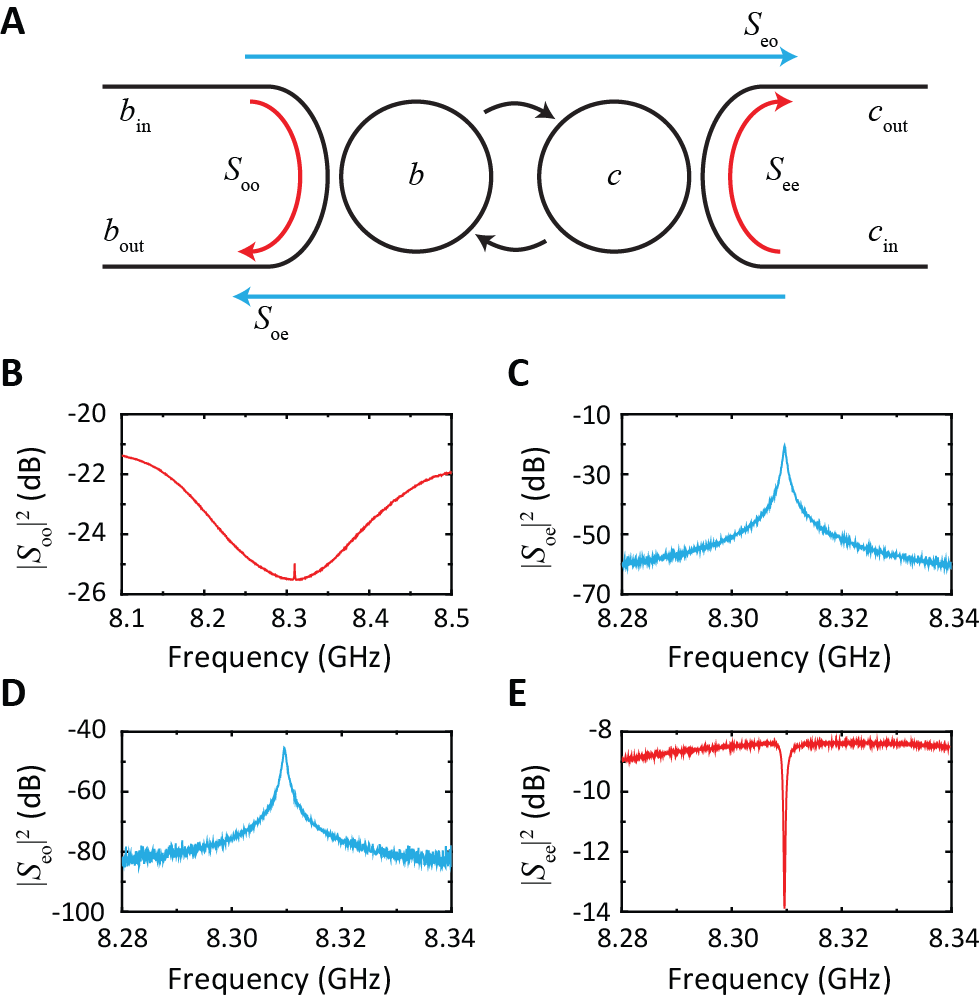}
\par\end{centering}
\caption{\textbf{Bidirectional Frequency Conversion. }(\textbf{A}) Schematic
showing the full conversion process. The optical reflection $S_{\mathrm{oo}}$
(\textbf{B}), microwave-to-optical conversion $S_{\mathrm{oe}}$ (\textbf{C}),
optical-to-microwave conversion $S_{\mathrm{eo}}$ (\textbf{D}), and
microwave reflection $S_{\mathrm{ee}}$ (\textbf{E})\textbf{ }are
measured in order to calibrate the on-chip conversion efficiency.
The control light power is 8$\,$dBm, and the DC voltage is 297$\,$V.
All conversion matrix coefficients are normalized to the RF output
power of the network analyzer (Supplementary Section VI).}

\label{Fig4}
\end{figure}

The bidirectional conversion is characterized by measuring the complete
conversion matrix (Fig.$\,$\ref{Fig4}A) with the DC voltage fixed
at $297\,\mathrm{V}$. By injecting the optical probe and monitoring
the microwave output, we measure the microwave-to-optical coefficient
$S_{\mathrm{oe}}$, which is a Lorentzian lineshape centered at the
microwave resonant frequency (Fig.$\,$\ref{Fig4}C). The optical-to-microwave
coefficient $S_{\mathrm{eo}}$ is measured by reversing the input
and output signal, which has the same spectrum shape with microwave-to-optical
coefficient $S_{\mathrm{oe}}$, indicating that the conversion is
bidirectional (Fig.$\,$\ref{Fig4}D). The optical and microwave reflection
spectra are also measured for calibration (Fig.$\,$\ref{Fig4}B \&
E), and the on-chip efficiency is estimated to be $(2.05\pm0.04)\%$
(Supplementary Section VIII). The main noise source during the conversion
process is the thermal excitation of the microwave cavity, which can
be reduced by working at lower temperature. And the noise generated
by the parametric amplification process is negligible because of the
deep resolved sideband condition (Supplementary Section IX).

\vbox{}

\noindent\textbf{\large{}Discussion}{\large \par}

An ideal quantum transducer demands the coherent conversion efficiency
approaching $100\%$, when the cooperativity equals unity and both
microwave and optical modes are deeply over-coupled. By optimizing
the fabrication process and material properties, the efficiency of
our device can be further increased. For instance, optical quality
factors above 2 million have been demonstrated with single crystalline
AlN \citep{liu2017aluminum}, therefore intrinsic loss for optical
modes can be reduced to $\kappa_{b,i}=2\pi\times100\:\mathrm{MHz}$.
Then, pump photon number can be increased by \textasciitilde{}100
times with the same pump power, and the enhanced coupling rate $g_{\mathrm{eo}}$
can reach $16\:\mathrm{MHz}$. With an improved microwave intrinsic
loss rate of $\kappa_{c,i}=2\pi\times10\:\mathrm{kHz}$ \citep{bruno2015reducing},
for example by using sapphire substrate, the optimal on-chip efficiency
exceeding $\eta=92\%$ can be achieved by choosing the external optical
coupling rate $\kappa_{b,\mathrm{ex}}=2\pi\times2.9\:\text{\ensuremath{\mathrm{GHz}}}$
and microwave coupling rate $\kappa_{c,\mathrm{ex}}=2\pi\times0.29\:\mathrm{MHz}$.
Therefore, with future development of single crystalline AlN on sapphire
system and its adaption for superconducting resonators, the approach
presented here is promising to realize high fidelity quantum state
transduction between superconducting and photonic circuits. 

\vbox{}

\noindent\textbf{\large{}Conclusion}{\large \par}

We have demonstrated the coherent photon transduction between integrated
superconducting and photonic circuits. High transduction efficiency
is realized based on triple-resonance electro-optics principle. Besides
high efficiency, low noise, and large bandwidth, the large frequency
tunability makes it easy to interface with different quantum systems,
and the planar structure allows the circuit-level integration of different
quantum devices. All these features not only make our device an ideal
quantum transducer, but also provide a scalable platform to synthesize
different quantum systems, paving the route towards large scale hybrid
quantum networks.

\vbox{}

\noindent\textbf{\large{}Materials and Methods}{\large \par}

\noindent\textbf{Polycrystal AlN film preparation} The 800$\,$nm
AlN film is grown on a Si wafer with 2-$\mathrm{\mu m}$-thick $\mathrm{SiO_{2}}$
layer by the radio frequency magnetron reactive sputtering, using
pure aluminum (99.999\%) targets in an argon and nitrogen gas mixture.
The sputtered AlN film is poly-crystal with $c$-axis highly oriented
perpendicular to the substrate. As AlN has wurtzite crystal structure,
the electro-optic coefficient has no direction dependence in the plain
perpendicular to $c$-axis. Thus the $r_{13}$ coefficient in the
sputtered AlN film can be used without considering the in-plain crystal
direction change \citep{xiong2012aluminum}. 

\noindent\textbf{NbTiN film preparation} NbTiN films are sputtered
by the RF magnetron method at room temperature using 70\% Nb and 30\%
Ti alloy target in an argon and nitrogen gas mixture. Amouphous NbTiN
film is formed uniformly on the surface, and the superconducting critical
temperature $T_{c}$ around $13.8\,\mathrm{K}$ is achieved.

\vbox{}

\noindent\textbf{\large{}Supplementary Materials}{\large \par}

\noindent Theory of cavity electro-optics and its utility for microwave-to-optical
conversion

\noindent Device fabrication procedure

\noindent Identifying phase matching conditions for optical modes

\noindent Influence of optical mode mixing on the vacuum coupling
rate $g_{\mathrm{eo}}$

\noindent Microwave resonator design

\noindent Measurement setup

\noindent Device temperature calibration

\noindent Efficiency calibration

\noindent Added noise during conversion

\noindent Fig.$\,$S1. Calculated internal conversion efficienty

\noindent Fig.$\,$S1. SEM picture of the cross section of a superconducting
cavity electro-optic device

\noindent Fig.$\,$S3. Anti-crossing between TE and TM optical modes.

\noindent Fig.$\,$S4. Measured spectrum signature of mixing between
TE and TM modes

\noindent Fig.$\,$S5. Wavelength dierence between adjacent resonances

\noindent Fig.$\,$S6. Phase matching wavelength and anti-crossing
strength

\noindent Fig.$\,$S7. Vacuum coupling rate with hybrid optical modes

\noindent Fig.$\,$S8. Microwave resonator simulation

\noindent Fig.$\,$S9. Experiment setup for microwave-to-optical photon
conversion

\noindent Fig.$\,$S10. Microwave resonator performance under different
temperature

\vbox{}

\noindent\textbf{\large{}References and Notes}{\large \par}

\vbox{}

\noindent\textbf{Acknowledgments} We acknowledge funding support
from an LPS/ARO grant (W911NF-14-1-0563), AFOSR MURI grant (FA9550-15-1-0029),
NSF EFRI grant (EFMA-1640959) and DARPA ORCHID program through a grant
from the Air Force Office of Scientific Research (AFOSR FA9550-10-1-0297),
and the Packard Foundation. Facilities used were supported by Yale
Institute for Nanoscience and Quantum Engineering and NSF MRSEC DMR
1119826. The authors thank L. Jiang for discussion, and M. Power,
M. Rooks and L. Frunzio for assistance in device fabrication.\textbf{
Author contributions} H.X.T., L.F., C.-L.Z. \& X.H. conceived the
experiment; L.F. fabricated the device; L.F., R.C., X.G., Z.G., \&
S.W. performed the experiment; L.F. \& C.-L.Z. analyzed the data.
L.F. \& C.-L.Z. wrote the manuscript, and all authors contribute to
the manuscript. H.X.T. supervised the work.\textbf{ Competing interests}
The authors declare that they have no competing interests.\textbf{
Data and materials availability} All data needed to evaluate the conclusions
in the paper are present in the paper and/or the Supplementary Materials.
Additional data related to this paper may be requested from the authors.

\clearpage{}

\newpage{}

\newpage{}

\onecolumngrid
\renewcommand{\thefigure}{S\arabic{figure}}
\setcounter{figure}{0} 
\renewcommand{\thepage}{S\arabic{page}}
\setcounter{page}{1} 
\renewcommand{\theequation}{S.\arabic{equation}}
\setcounter{equation}{0} 
\setcounter{section}{0}

\part*{\textsc{\LARGE{}Supplementary Information}}

\author{Linran Fan, Chang-Ling Zou, Risheng Cheng, Xiang Guo, Xu Han, Zheng
Gong, Sihao Wang}

\affiliation{Department of Electrical Engineering, Yale University, New Haven,
Connecticut 06511, USA}

\author{Hong X. Tang}
\email{hong.tang@yale.edu}

\selectlanguage{english}%

\affiliation{Department of Electrical Engineering, Yale University, New Haven,
Connecticut 06511, USA}

\tableofcontents{}

\section{Theory of cavity electro-optics and its utility for microwave-to-optical
conversion}

The electro-optic effect is a second-order nonlinear process {[}30{]},
thus the interaction Hamiltonian can be written as 
\begin{equation}
H_{\mathrm{I}}=\hbar g_{\mathrm{eo}}(a+a^{\dagger})(b+b^{\dagger})(c+c^{\dagger})\label{eq:S1.1-1}
\end{equation}
where $a$, $b$, and $c$ are the Bosonic annihilation operators
for the two optical modes and microwave mode respectively, and $g_{\mathrm{eo}}$
is the vacuum electro-optic coupling strength. As the optical frequency
is much higher than the microwave frequency and $g_{\mathrm{eo}}$,
the counter rotating terms $a^{\dagger}b^{\dagger}+ab$ is neglected.
Thus, the interaction Hamiltonian becomes
\begin{equation}
H_{\mathrm{I}}=\hbar g_{\mathrm{eo}}(ab^{\dagger}+a^{\dagger}b)(c+c^{\dagger})\label{eq:S1.1-1-1}
\end{equation}
And the vacuum coupling strength $g_{\mathrm{eo}}$ can be expressed
as 
\begin{equation}
\hbar g_{\mathrm{eo}}=-\frac{\int(\varepsilon_{a,i}\varepsilon_{b,j}r_{ijk})\cdot(u_{a,i}u_{b,j}^{*}u_{c,k})dxdydz}{8\pi\sqrt{\varepsilon_{0}}\prod_{l=a,b,c}\sqrt{\int\varepsilon_{l,i}u_{l,i}^{*}u_{l,i}dxdydz/\hbar\omega_{l}}}.\label{eq:S1.2-1}
\end{equation}
Here $\varepsilon_{l,i}$ and $u_{l,i}$ ($l\in\{a,b,c\}$, $i\in\{x,y,z\}$)
denote relative permittivity and electric field components respectively,
$\omega_{l}$ is the angular frequency, $r_{ijk}$ is the electro-optic
component, and Einstein summation convention is used.

If we consider a ring structure with radius $R$, the field distribution
of optical whispering gallery modes and microwave modes in the cylindrical
coordinator can be expressed as $u_{l,i}\left(r,z,\theta\right)=u_{l,\perp,i}\left(r,z\right)e^{-im_{l}\theta}$
with $l\in\{a,b\}$ and $u_{c,i}\left(r,z,\theta\right)=\sum_{m_{c}}x_{c,m_{c}}u_{c,m_{c},\perp,i}\left(r,z\right)e^{-im_{c}\theta}$,
respectively, where $m_{l}$ is the azimuthal number, and $x_{c,m}$
indicates the contribution of different azimuthal numbers for microwave
modes. Thus the vacuum coupling rate is
\begin{equation}
\hbar g_{\mathrm{eo}}=\sqrt{\frac{1}{2\pi\varepsilon_{0}R}}\frac{\int(\varepsilon_{a,i}\varepsilon_{b,j}r_{ijk})\cdot(u_{a,\perp,i}u_{b,\perp,j}^{*}u_{c,m_{c},\perp,k})drdz}{8\pi\prod_{l=a,b,c}\sqrt{\int\varepsilon_{l,i}u_{l,\perp,i}^{*}u_{l,\perp,i}drdz/\hbar\omega_{l}}}\times x_{c,m_{c}}.\label{eq:S1.2-1-1}
\end{equation}
with $m_{c}=m_{b}-m_{a}$. The expression indicates that the microwave
field should have non-zero coefficient for the azimuthal number $m_{c}=m_{b}-m_{a}$.

If mode $a$ is coherently driven with a strong pump, the system Hamiltonian
in the resolved sideband regime will become
\begin{equation}
H=\hbar\omega_{a}a^{\dagger}a+\hbar\omega_{b}b^{\dagger}b+\hbar\omega_{c}c^{\dagger}c+H_{\mathrm{I}}+i\sqrt{\frac{\kappa_{a,\mathrm{ex}}P_{a}}{\hbar\omega_{\mathrm{p}}}}(a^{\dagger}e^{-i\omega_{\mathrm{p}}t}-ae^{i\omega_{\mathrm{p}}t})\label{eq:S1.3-1}
\end{equation}
where $P_{a}$ , $\omega_{\mathrm{p}}$, and $\kappa_{a,\mathrm{ex}}$
are the pump power, pump frequency, and external coupling rate of
mode $a$, respectively. For a strong coherent pump field, mode $a$
is in steady state and the backaction from the system to the pump
mode is negligible, thus we can treat the pump mode as a complex number

\begin{equation}
\left\langle a\right\rangle =\sqrt{\frac{\kappa_{a,\mathrm{ex}}}{\kappa_{a}^{2}/4+\delta_{a}^{2}}}\times\sqrt{\frac{P_{a}}{\hbar\omega_{\mathrm{p}}}}e^{i\phi},\label{eq:S1.4-1}
\end{equation}
with the corresponding pump cavity photon number 
\begin{equation}
n_{a}=\left\langle a^{\dagger}a\right\rangle =\left|\left\langle a\right\rangle \right|^{2}
\end{equation}
Here $\delta_{a}=\omega_{a}-\omega_{\mathrm{p}}$ is the pump detuning,
$\kappa_{a}=\kappa_{a,\mathrm{ex}}+\kappa_{a,\mathrm{i}}$ is the
total decay rate, $\kappa_{a,\mathrm{i}}$ is the intrinsic loss rate
of mode $a$, and $\phi$ is the pump phase. Choosing the pump phase
$\phi=0$ and the pump mode frequency close to $\omega_{b}-\omega_{c}$,
the system Hamiltonian in the rotation frame of $\hbar\omega_{\mathrm{p}}b^{\dagger}b$
under rotating-wave approximation (RWA) can be simplified to 
\begin{equation}
H=\hbar\delta_{b}b^{\dagger}b+\hbar\omega_{c}c^{\dagger}c+\hbar G(b^{\dagger}c+bc^{\dagger})\label{eq:S1.5}
\end{equation}
with $G=\left\langle a\right\rangle g_{\mathrm{eo}}=\sqrt{n_{a}}g_{\mathrm{eo}}$,
which is the enhanced coupling rate, and $\delta_{b}=\omega_{b}-\omega_{\mathrm{p}}$.
Equation$\,$(\ref{eq:S1.5}) has the form of a linear beam splitter
Hamiltonian, allows the coherent state transfer between mode $b$
and mode $c$, which has the potential for quantum state transfer
between microwave and optical frequencies. The equation of motion
including the input and output of mode $b$ and $c$ can be written
as 
\begin{align}
 & \frac{d}{dt}b=-(i\delta_{b}+\frac{\kappa_{b}}{2})b-iGc+\sqrt{\kappa_{b,\mathrm{ex}}}b_{\mathrm{in}}e^{-i\delta_{b,\mathrm{in}}t}\label{eq:S1.7}\\
 & \frac{d}{dt}c=-(i\omega_{c}+\frac{\kappa_{c}}{2})c-iGb+\sqrt{\kappa_{c,\mathrm{ex}}}c_{\mathrm{in}}e^{-i\omega_{c,\mathrm{in}}t}\label{eq:S1.8}\\
 & b_{\mathrm{out}}=b_{\mathrm{in}}-\sqrt{\kappa_{b,\mathrm{ex}}}b\label{eq:S1.9}\\
 & c_{\mathrm{out}}=c_{\mathrm{in}}-\sqrt{\kappa_{c,\mathrm{ex}}}c\label{eq:S1.10}
\end{align}
where $b_{\mathrm{in}}$ and $c_{\mathrm{in}}$ denote the input signals,
$\delta_{b,\mathrm{in}}=\omega_{b,\mathrm{in}}-\omega_{\mathrm{p}}$
and $\omega_{c,\mathrm{in}}$ are the angular frequency of inputs,
$\kappa_{b}$ and $\kappa_{c}$ are the total decay rates, $\kappa_{b,\mathrm{ex}}$
and $\kappa_{c,\mathrm{ex}}$ represent the external coupling rate
of mode $b$ and $c$, respectively. Solving Eqs.$\:$(\ref{eq:S1.7})$\:$-$\:$(\ref{eq:S1.10})
in steady state, we can get the relationship between the input and
output of mode $b$ and $c$ as
\begin{equation}
\begin{pmatrix}b_{\mathrm{out}}\\
c_{\mathrm{out}}
\end{pmatrix}=\begin{pmatrix}1-\frac{\kappa_{b,\mathrm{ex}}}{i(\delta_{b}-\delta_{b,\mathrm{in}})+\frac{\kappa_{b}}{2}+\frac{G^{2}}{i(\omega_{c}-\delta_{b,\mathrm{in}})+\frac{\kappa_{c}}{2}}} & \frac{-iG\sqrt{\kappa_{b,\mathrm{ex}}\kappa_{c,\mathrm{ex}}}}{G^{2}+[i(\omega_{c}-\omega_{c,\mathrm{in}})+\frac{\kappa_{c}}{2}][i(\delta_{b}-\omega_{c,\mathrm{in}})+\frac{\kappa_{b}}{2}]}\\
\frac{-iG\sqrt{\kappa_{b,\mathrm{ex}}\kappa_{c,\mathrm{ex}}}}{G^{2}+[i(\delta_{b}-\delta_{b,\mathrm{in}})+\frac{\kappa_{b}}{2}][i(\omega_{c}-\delta_{b,\mathrm{in}})+\frac{\kappa_{c}}{2}]} & 1-\frac{\kappa_{c,\mathrm{ex}}}{i(\omega_{c}-\omega_{c,\mathrm{in}})+\frac{\kappa_{c}}{2}+\frac{G^{2}}{i(\delta_{b}-\omega_{c,\mathrm{in}})+\frac{\kappa_{b}}{2}}}
\end{pmatrix}\begin{pmatrix}b_{\mathrm{in}}\\
c_{\mathrm{in}}
\end{pmatrix}.\label{eq:S1.11}
\end{equation}

\begin{figure}
\begin{centering}
\includegraphics{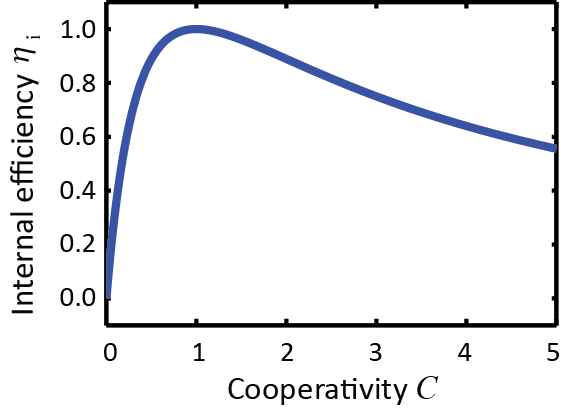}
\par\end{centering}
\caption{\textbf{Calculated internal conversion efficiency.}}

\label{FigS1.1}
\end{figure}

Thus, the bidirectional conversion efficiency can be written as 
\begin{equation}
\eta=|\frac{b_{\mathrm{out}}}{c_{\mathrm{in}}}|^{2}=|\frac{c_{\mathrm{out}}}{b_{\mathrm{in}}}|^{2}=\frac{\kappa_{b,\mathrm{ex}}}{\kappa_{b}}\frac{\kappa_{c,\mathrm{ex}}}{\kappa_{c}}\frac{4C}{|C+(1+\frac{2i(\delta_{b}-\omega)}{\kappa_{b}})(1+\frac{2i(\omega_{c}-\omega)}{\kappa_{c}})|^{2}}\label{eq:S1.12}
\end{equation}
where $C=\frac{4G^{2}}{\kappa_{b}\kappa_{c}}$ is defined as the system
cooperativity, and $\omega$ is the input signal angular frequency.
The maximum conversion is reached when $\omega=\omega_{c}$ and the
pump frequency matches the frequency difference between mode $b$
and $c$ ($\omega_{\mathrm{p}}=\omega_{b}-\omega_{c}$), and the on-chip
conversion efficiency is 
\begin{equation}
\eta=\frac{\kappa_{b,\mathrm{ex}}}{\kappa_{b}}\frac{\kappa_{c,\mathrm{ex}}}{\kappa_{c}}\times\frac{4C}{(1+C)^{2}}.\label{eq:S1.13}
\end{equation}
with conversion bandwidth $(1+C)\kappa_{c}$. As we can see from Eq.$\,$(\ref{eq:S1.13}),
the conversion efficiency consists of two parts: the cavity extraction
efficiency which is $\frac{\kappa_{b,\mathrm{ex}}}{\kappa_{b}}\frac{\kappa_{c,\mathrm{ex}}}{\kappa_{c}}$,
and the internal conversion efficiency which is 
\begin{equation}
\eta_{\mathrm{i}}=\frac{4C}{(1+C)^{2}}.\label{eq:S1.14}
\end{equation}
Figure$\:$\ref{FigS1.1} plots the achievable internal conversion
efficiency dependence on the cooperativity. The maximum internal conversion
efficiency ($\eta_{\mathrm{i}}=1$) is achieved when the cooperativity
$C=1$. Also the internal conversion efficiency can never be larger
than unity, due to the photon number conservation imposed by Eq.$\,$(\ref{eq:S1.5}).

\section{Device fabrication procedure}

\begin{figure}
\begin{centering}
\includegraphics{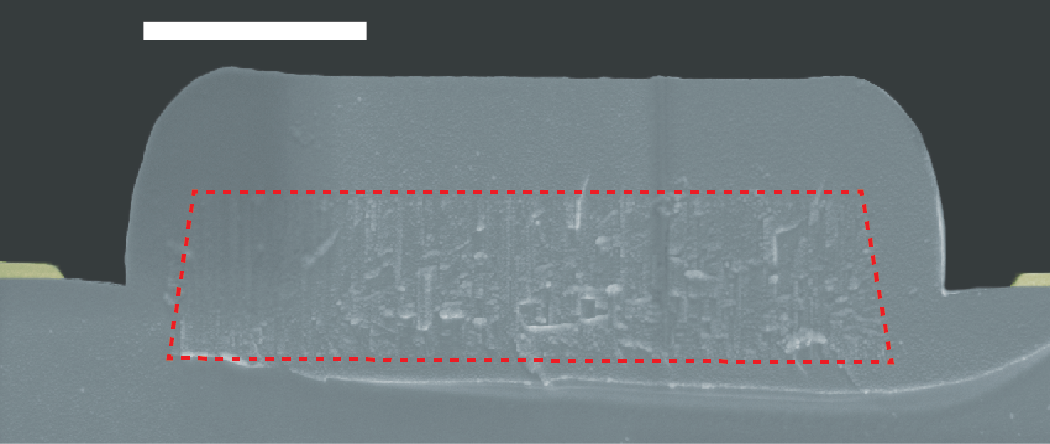}
\par\end{centering}
\caption{\textbf{SEM picture of the cross section of a superconducting cavity
electro-optic device. AlN (red) is covered with SiO$_{2}$ (grey),
and NbTiN (yellow) is sputtered on top of SiO$_{2}$. The device shown
in the SEM picture has a different waveguide width ($3\,\mathrm{\mu m}$)
with the one used for experiment ($2\,\mathrm{\mu m}$). The scale
bar is $1\,\mathrm{\mu m}$.}}

\label{FigS5.1-1}
\end{figure}

The 800$\,$nm AlN film is grown on a Si wafer with 2-$\mathrm{\mu m}$-thick
$\mathrm{SiO_{2}}$ layer by the radio frequency magnetron reactive
sputtering, using pure aluminum (99.999\%) targets in an argon and
nitrogen gas mixture. Optical waveguides are patterned with electron
beam lithography (EBL) using hydrogen silsesquioxane (HSQ) resist,
subsequently transferred to the AlN layer by chlorine-based reactive
ion-etching (RIE). Then 400$\,$nm $\mathrm{SiO_{2}}$ is deposited
with plasma-enhanced chemical vapor deposition (PECVD), followed by
device annealing at $940\,^{\mathrm{o}}\mathrm{C}$. After annealing,
NbTiN superconducting film is sputtered on top of SiO\textsubscript{2}by
the magnetron sputtering method. A second EBL is performed to define
the microwave resonator pattern with HSQ, which is transferred to
the NbTiN layer by chlorine-based RIE. A test device instead of the
device used in experiment is cleaved along with the diameter of the
AlN optical ring resonator to verify the consistence beween fabricated
device and our design. The SEM picture of the device cross section
is shown in Fig. \ref{FigS5.1-1}. As we can see from the SEM picture,
the optical waveguide width ($3\,\mathrm{\mu m}$) and the gap between
optical waveguide and microwave electrode ($400\,\mathrm{nm}$) match
our design.

\section{Identifying phase matching conditions for optical modes}

\begin{figure}
\begin{centering}
\includegraphics{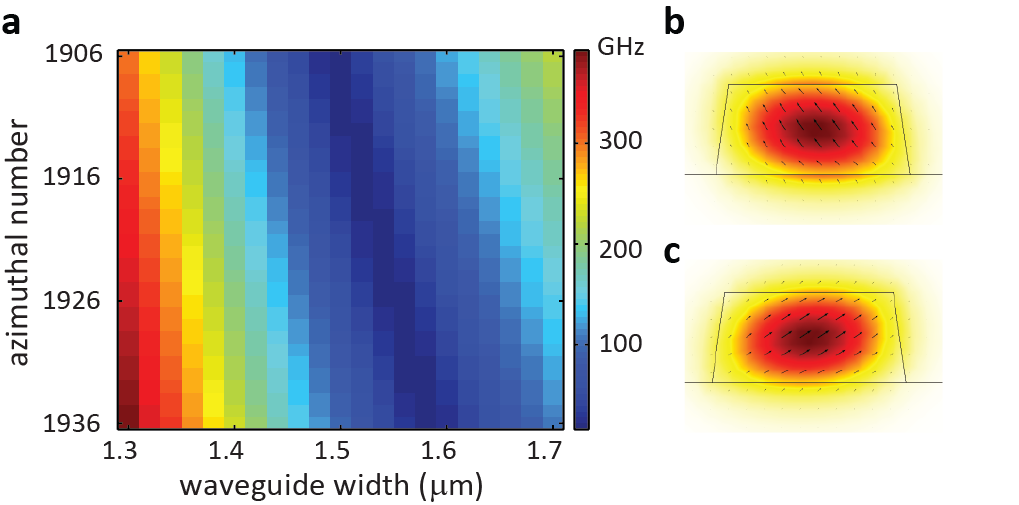}
\par\end{centering}
\caption{\textbf{Anti-crossing between TE and TM optical modes.} \textbf{a}
Simulated frequency difference between TE and TM optical modes with
the same azimuthal number ($m_{a}=m_{b}$) around 194$\:$THz frequency.
\textbf{b, c} Mode profiles of the hybrid optical modes when the frequency
difference is smallest in \textbf{a}. Arrows represent the electric
field direction, and the color saturation in \textbf{b} and \textbf{c}
shows the mode energy density. Ring radius is 240$\,\mu$m, and the
AlN thickness is 800$\:$nm in simulation.}

\label{FigS2.1}
\end{figure}

We use the $r_{13}$ electro-optic coefficient of AlN, therefore the
fundamental TE optical mode, fundamental TM optical mode, and the
TE microwave mode are needed in our triple-resonance scheme. As the
microwave mode has much smaller frequency and azimuthal number than
optical modes, we first identify the condition that TE and TM optical
modes have the same frequency and azimuthal number. The waveguide
geometry is designed that TE and TM modes have the same phase velocity,
thus the same azimuthal number. In a ring structure, the frequency
degeneracy between TE and TM optical modes with the same azimuthal
number will be broken due to the non-vertical sidewall of AlN waveguides
as shown in Fig.$\:$\ref{FigS2.1}a. Anti-crossing between TE and
TM modes can be observed, with the mode coupling strength (half of
the minimum frequency difference) determined by the sidewall angle
and the ring radius. At the anti-crossing point, the two optical modes
cannot be classified as pure TE or TM mode anymore. Instead, the two
optical modes are the mixture of TE and TM modes (Fig.$\:$\ref{FigS2.1}b
and c).

The mixing between TE and TM modes provides the spectrum signature
to identify the phase matching condition. If we tune the optical input
in the bus waveguide to be TE mode, we can only observe one group
of TE optical modes at the output if there is no mixing between TE
and TM modes. If the TE and TM azimuthal numbers approach each other,
mixing between TE and TM modes takes place, and we can observe two
groups of modes, corresponding to the two mixed modes. 

\begin{figure}
\begin{centering}
\includegraphics{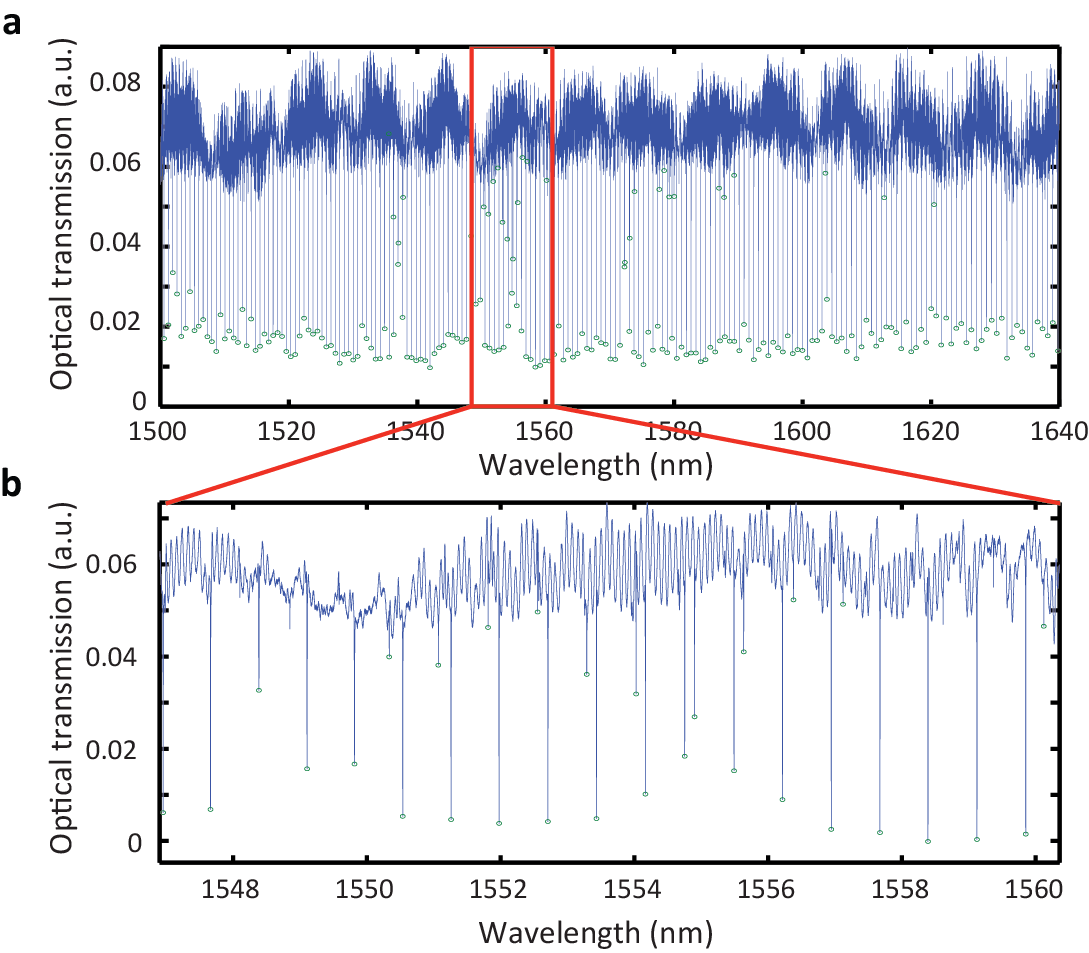}
\par\end{centering}
\caption{\textbf{Measured spectrum signature of mixing between TE and TM modes.}
\textbf{a} Optical transmission spectrum with pure TE optical input.
\textbf{b} Zoom-in of the optical transmission spectrum around 1555$\:$nm.
Mode splitting and extinction decrease indicate the mixing between
TE and TM modes. }

\label{FigS2.2}
\end{figure}

\begin{figure}
\begin{centering}
\includegraphics{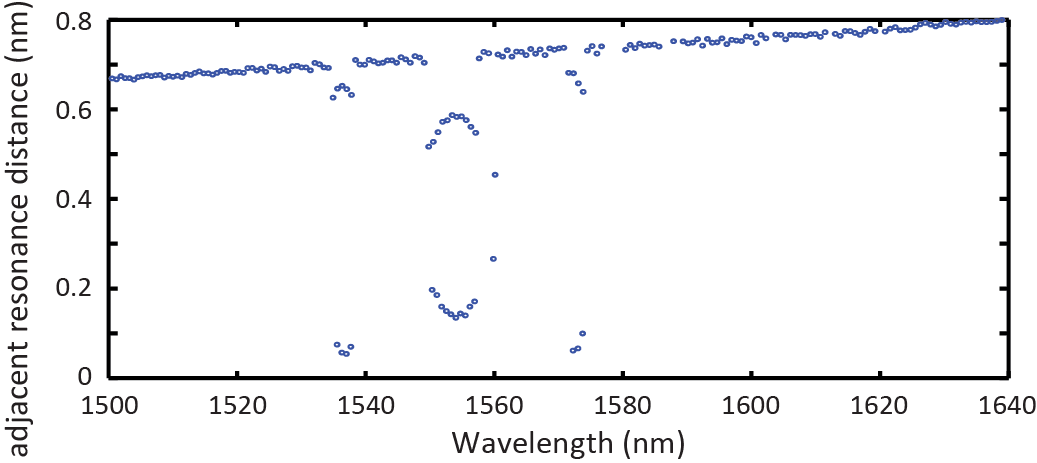}
\par\end{centering}
\caption{\textbf{Wavelength difference between adjacent resonances.} The resonant
wavelengths of all optical modes from Fig.$\:$\ref{FigS2.2} are
extracted, and the adjacent resonance distance is computed by subtracting
the resonant wavelength of the one optical mode by its previous optical
mode.}

\label{FigS2.3}
\end{figure}

\begin{figure}
\begin{centering}
\includegraphics{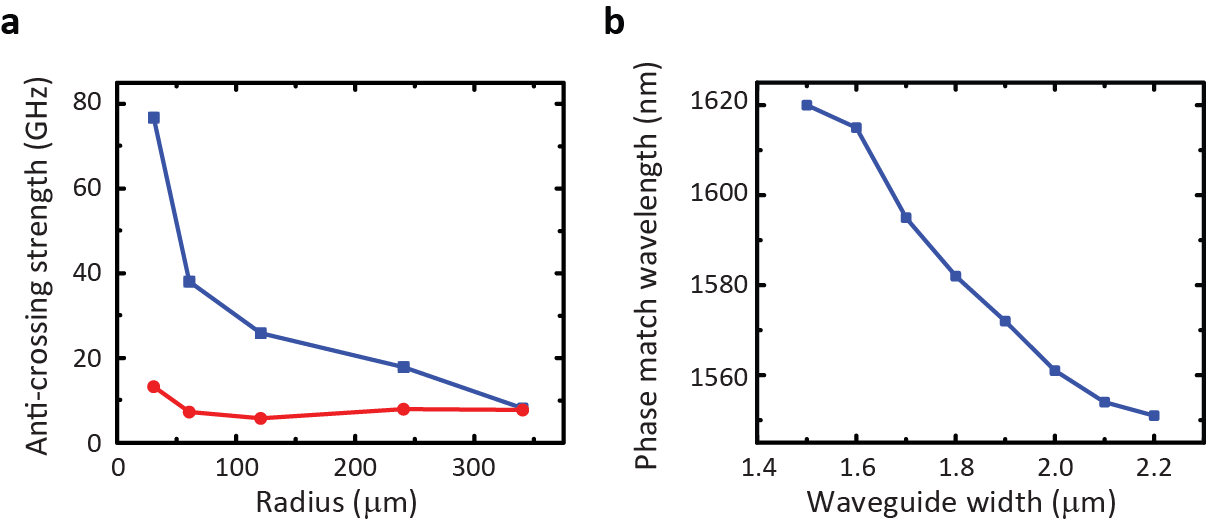}
\par\end{centering}
\caption{\textbf{Phase matching wavelength and anti-crossing strength.} \textbf{a}
Anti-crossing strength dependence on ring radius. The cases for $m_{a}=m_{b}$
and $|m_{a}-m_{b}|=1$ are shown in blue and red respectively. \textbf{b
}Phase matching wavelength dependence on the waveguide width, with
ring radius 240$\:\mu$m. }

\label{FigS2.4}
\end{figure}

In Fig.$\:$\ref{FigS2.2}, we show the measured TE optical spectrum
of a AlN ring with radius 240$\:\mu$m and width 2.1$\:\mu$m. Beside
the fundamental TE optical modes, we observe another group of modes
around 1555$\:$nm, showing the mixing between TE and TM modes. In
addition, the mode extinction ratio also drops, as the coupling rate
between the bus waveguide and ring resonator drops. In Fig.$\:$\ref{FigS2.3},
we plot the wavelength difference between two adjacent optical resonances.
When there is no mixing between TE and TM modes, the wavelength difference
corresponds to the free spectral range of the ring resonator. When
the mixing happens, the wavelength difference drops below half of
the free spectral range, and the smallest wavelength difference gives
us the anti-crossing strength between TE and TM modes. Beside the
main anti-crossing around 1555$\,$nm, we can also observe anti-crossing
around 1538$\,$nm and 1573$\,$nm, corresponding to the azimuthal
number different of $1$. As we can see, the anti-crossing strength
is much smaller for $|m_{a}-m_{b}|=1$ compared with $m_{a}=m_{b}$
(Fig.$\:$\ref{FigS2.4}a). Here, the mode mixing between the TE and
TM modes for $|m_{a}-m_{b}|=1$ is from the surface roughness and
the perturbation induced by the external coupling waveguide.

As the vacuum coupling efficiency is inversely proportional to the
square root of the ring radius (Eq.$\:$(\ref{eq:S1.2-1})), we should
minimize the ring radius to maximize the vacuum coupling rate. Therefore,
instead of using TE and TM modes with the same azimuthal number ($m_{a}=m_{b}$),
we use the TE and TM modes with azimuthal number different by 1 ($|m_{a}-m_{b}|=1$),
which have a much smaller mode anti-crossing. Accordingly, the microwave
azimuthal number should also be designed to be $1$, which will be
explained in detail in Sect. III.

With the same ring radius, the phase matching wavelength can be fine-tuned
by the waveguide width, which modifies the phase velocity differently
for the TE and TM optical modes. Figure$\:$\ref{FigS2.4}b shows
that the phase matching wavelength can be tuned by 70$\:$nm with
a waveguide width change of 600$\:$nm. Therefore we can precisely
control the phase matching wavelength to match different wavelength
bands, providing the possibility for wavelength domain multiplexing.

\section{Influence of optical mode mixing on the vacuum coupling rate $g_{\mathrm{eo}}$}

\begin{figure}
\begin{centering}
\includegraphics{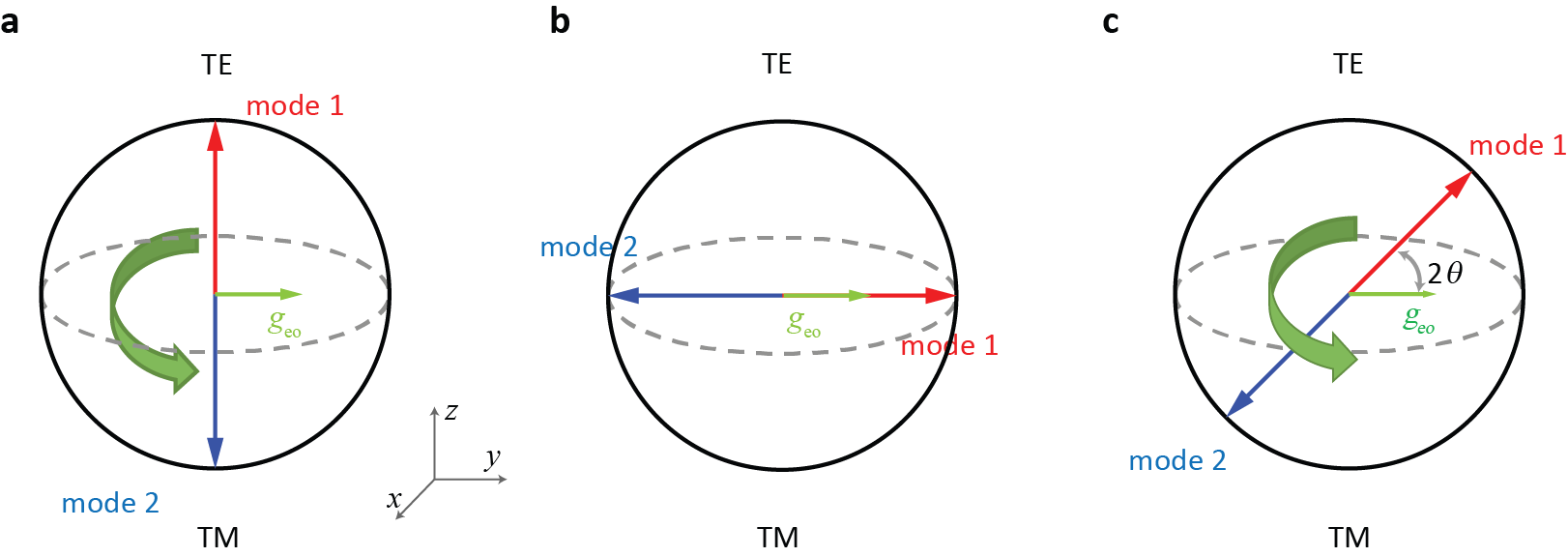}
\par\end{centering}
\caption{\textbf{Vacuum coupling rate with hybrid optical modes.} In Bloch
sphere representation, the north and south poles are the TE and TM
optical modes respectively, and the electro-optic interaction is equivalent
to the rotation along $y$-axis. Three cases are shown: \textbf{a,}
no mixing; \textbf{b,} maximum mixing; \textbf{c} partial mixing.
The effective coupling rate will be $g$, 0, and $g\cos2\theta$ respectively.}

\label{FigS3.1}
\end{figure}

As we can see from the last section, the anti-crossing between TE
and TM optical modes sets the lowest possible microwave frequency
if triple resonance scheme is used. Furthermore, the vacuum coupling
rate is decreased due to the mixing between TE and TM optical modes.
The system Hamiltonian for the mode mixing is 
\begin{equation}
H_{\mathrm{x}}=\hbar g_{\mathrm{\mathrm{x}}}(a^{\dagger}b+ab^{\dagger}).\label{eq:S3.1-1}
\end{equation}
where $g_{\mathrm{x}}$ is the mode mixing strength between TE and
TM optical modes. Since $g_{\mathrm{x}}\gg g_{\mathrm{eo}}$, we first
diagonalize the optical part of the Hamiltonian by introducing the
hybrid optical modes as 
\begin{equation}
\begin{pmatrix}A\\
B
\end{pmatrix}=\begin{pmatrix}\cos\theta & \sin\theta\\
-\sin\theta & \cos\theta
\end{pmatrix}\begin{pmatrix}a\\
b
\end{pmatrix}\label{eq:S3.2}
\end{equation}
where $A$ and $B$ are the operators for hybrid optical modes, and
$\theta\in[-\frac{\pi}{4},\frac{\pi}{4}]$, satisfying 
\begin{equation}
\tan2\theta=\frac{2g_{\mathrm{x}}}{\omega_{a}-\omega_{b}}.\label{eq:S3.2-1}
\end{equation}
The eigenfrequencies of the hybrid optical modes are 
\begin{align}
\omega_{A} & =\frac{\omega_{a}+\omega_{b}}{2}+\sqrt{g_{\mathrm{x}}^{2}+(\frac{\omega_{a}-\omega_{b}}{2})^{2}},\label{eq:S3.3}\\
\omega_{B} & =\frac{\omega_{a}+\omega_{b}}{2}-\sqrt{g_{\mathrm{x}}^{2}+(\frac{\omega_{a}-\omega_{b}}{2})^{2}}.\label{eq:S3.4}
\end{align}
The minimum frequency difference is therefore $2g_{\mathrm{x}}$,
which corresponding to the frequency gap in the anti-crossing spectrum. 

By inverting Eq.$\:$(\ref{eq:S3.2}), we can obtain
\begin{equation}
\begin{pmatrix}a\\
b
\end{pmatrix}=\begin{pmatrix}\cos\theta & -\sin\theta\\
\sin\theta & \cos\theta
\end{pmatrix}\begin{pmatrix}A\\
B
\end{pmatrix}.\label{eq:S3.5}
\end{equation}
Plugging Eq.$\:$(\ref{eq:S3.5}) into Eq.$\:$(\ref{eq:S1.1-1-1}),
we obtain the modified electro-optic interaction Hamiltonian
\begin{align}
H_{\mathrm{I,m}} & =\hbar g_{\mathrm{eo}}\cos2\theta(A^{\dagger}B+AB^{\dagger})(c+c^{\dagger}).\label{eq:S3.6}\\
 & =\hbar g_{\mathrm{eo}}^{\mathrm{eff}}(A^{\dagger}B+AB^{\dagger})(c+c^{\dagger}).
\end{align}
Comparing with the original interaction Hamiltonian, the modified
Hamiltonian has the same form, but lower effective coupling strength
$g_{\mathrm{eo}}^{\mathrm{eff}}=g_{\mathrm{eo}}\cos2\theta$. 

When the detuning between the two original optical modes is zero,
thus $\theta=45^{\mathrm{o}}$, the effective electro-optic interaction
vanishes. This phenomena can be intuitively shown with the Bloch sphere.
Assuming the north and south poles are the TE and TM optical modes
respectively, then the electro-optic interaction is equivalent to
the rotation along $y$-axis (Fig.$\:$\ref{FigS3.1}a). When there
is mode anti-crossing and the detune is zero, the two renormalized
optical modes involved are maximally hybridized mode, whose directions
are along $y$-axis, thus the rotation along $y$-axis will not induce
the mode conversion between the two renormalized modes (Fig.$\:$\ref{FigS3.1}b).
Therefore, the detuning must be non-zero, and the effective vacuum
coupling rate is $g_{\mathrm{eo}}\cos2\theta$ (Fig.$\:$\ref{FigS3.1}c).
When the detuning is much larger than the anti-crossing strength,
the vacuum coupling rate approaches the original value $g_{\mathrm{eo}}$.

\section{Microwave resonator design}

\begin{figure}
\begin{centering}
\includegraphics{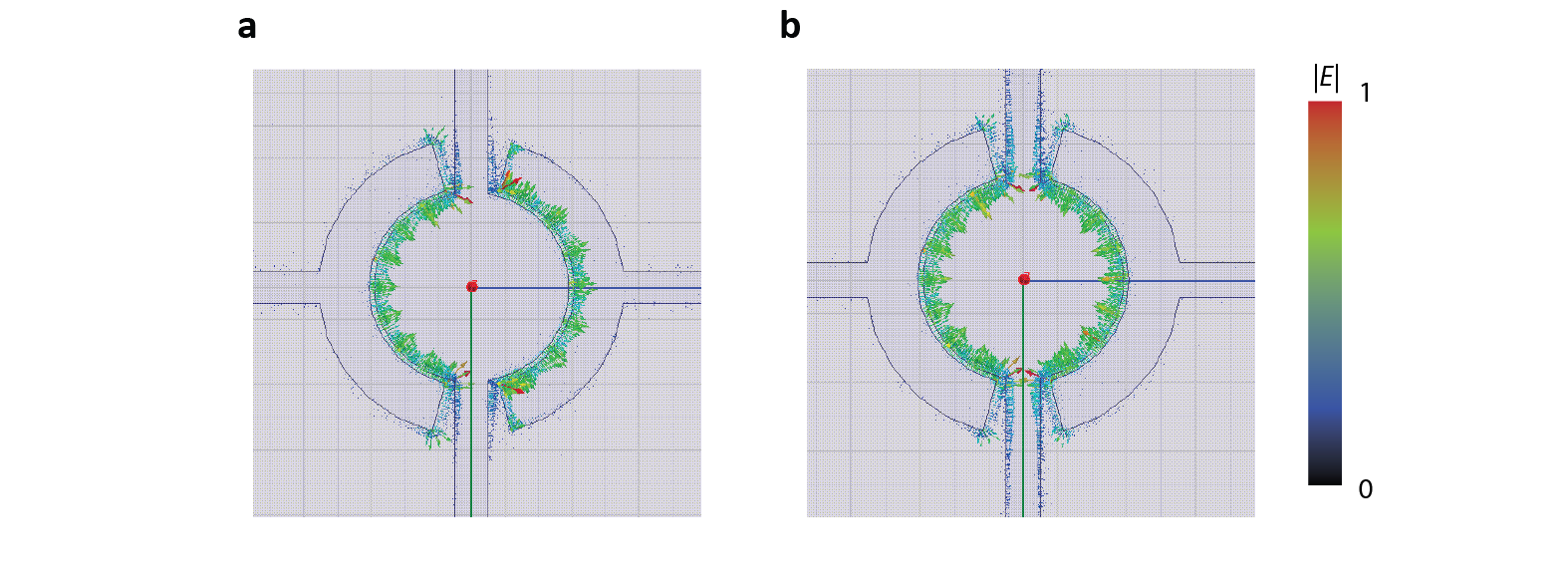}
\par\end{centering}
\caption{\textbf{Microwave resonator simulation.} \textbf{a} Simulated electric
field distribution of the anti-symmetric mode. \textbf{b} Simulated
electric field distribution of the symmetric mode. The microwave modes
shown here are simulated by ANSYS HFSS, and the arrow color shows
the relative strength of the elelctric field.}

\label{FigS4.1}
\end{figure}

This planar microwave resonator can be treated as two coupled lumped
element microwave resonators, where the central disk and surrounding
arms provide the dominant capacitance and inductance respectively.
The two lumped microwave resonators are coupled through the connection
at the central disk. The resonator has symmetric and anti-symmetric
modes due to the coupling between capacitors. The simulated electric
field distribution is plotted in Fig.$\:$\ref{FigS4.1}a and b. The
symmetric mode has uniformly distributed electric field with the azimuthal
angle $\phi$, thus the azimuthal number $m_{c}=0$, satisfying the
phase matching condition for $m_{a}=m_{b}$. And the anti-symmetric
mode has a $\sin\phi$ dependence, thus strong component of $m_{c}=1$,
satisfying the phase matching condition for $|m_{a}-m_{b}|=1$.

The lumped element microwave resonator can provide very small electric
mode volume, which is critical for enhancing the vacuum coupling rate
(Eq.$\:$\ref{eq:S1.2-1}). Also the resonant frequency can be varied
by adjusting the arm length of the resonator. Thus the capacitance
part of the resonator can be kept fixed, making it easy to match the
optical cavity and resonant frequency simultaneously. Another advantage
of this design is that the long-arm inductor allows supercurrents
to generate magnetic flux far-extended from the chip surface, making
it feasible to inductively couple the microwave resonator with an
off-chip loop probe for broadband high-efficiency microwave signal
input and readout.

\section{Measurement setup}

\begin{figure}
\begin{centering}
\includegraphics{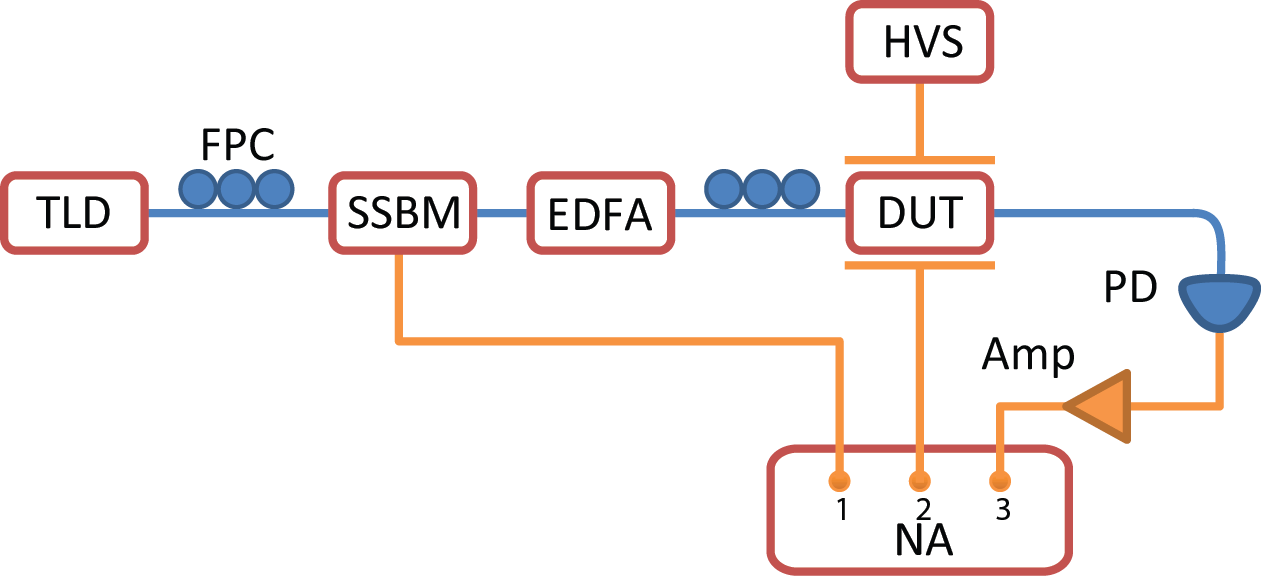}
\par\end{centering}
\caption{\textbf{Experiment setup for microwave-to-optical photon conversion.}
TLD, tunable laser diode; EDFA, erbium doped fiber amplifier; SSBM,
single side-band modulator; FPC, fiber polarization controller; DUT,
device under test; HVS, high voltage source; PD, photo-detector; Amp,
RF amplifier; NA network analyzer.}

\label{FigS5.1}
\end{figure}

The experiment setup is shown in Fig.$\:$\ref{FigS5.1}. Coherent
light from a tunable laser diode (TLD) is used as the control light,
which is sent into an single side-band modulator (SSBM) to generate
a weak probe light. The modulation RF signal is from a network analyzer
(NA). Then a erbium doped fiber amplifier (EDFA) is used to amplify
the control light. Light is coupled to and from the on-chip bus waveguide
through a pair of grating couplers, which only transmit TE light.
The output light is detected by the high frequency photodetector (PD).
The output signal of the photodetector is amplified by a RF amplifier
(Amp), and de-modulated with network analyzer. The microwave port
of the device is directly connect to the network analyzer. A high
voltage source (HVS) is used to provide the static electric field
across the device for optical resonance tuning.

\section{Device temperature calibration}

\begin{figure}
\begin{centering}
\includegraphics{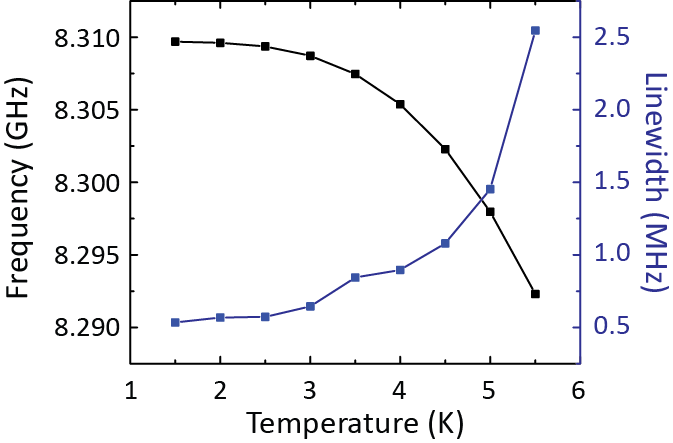}
\par\end{centering}
\caption{\textbf{Microwave resonator performance under different temperature.}}

\label{FigS6.1}
\end{figure}

The strong control light can heat up the device, which may degrade
the microwave resonator performance. In order to calibrated the device
temperature, we measure the microwave resonant frequency and linewidth
at different ambient temperatures without light input. As we can see
from Fig.$\:$\ref{FigS6.1}, the resonator frequency drops and the
linewidth increases with the temperature increase. Therefore, by measuring
the microwave resonant frequency and linewidth at base temperature
with strong control light, we can infer the effective temperature
of the microwave cavity. The linewidths at 2.0$\,$K, 2.5$\,$K, and
3.0$\,$K are 0.64$\,$MHz, 0.58$\,$MHz, and$\,$0.54 MHz respectively.
As shown in Fig.$\,$2E and Fig.4$\,$E in the main text, the microwave
resonance linewidth is around 0.55$\,$MHz under 8$\,$dBm optical
pump, indicating that the effective temperature of the device is around
2.1$\,$K.

\section{Efficiency calibration}

The cooperativity $C$ of $0.075\pm0.001$ is estimated from the fitting
of the blue curve in Fig.$\:$3C in the main text, leading to internal
efficiency $\eta_{\mathrm{i}}$ around $(25.9\pm0.3)\%$ based on
Eq.$\:$(\ref{eq:S1.14}). From the resonance extinction ratio $|R_{i}|^{2}$,
the photon extraction efficiency can be calculated $\eta_{i,\mathrm{ex}}=\frac{\kappa_{i,\mathrm{ex}}}{\kappa}=\frac{1-|R|}{2}$
assuming under-coupled condition with $i=b,\,c$ for optical signal
and microwave modes respectively. From Fig. 2$\:$C \& E in the main
text, the resonance extinction ratio of optical signal and microwave
modes can be extracted, $|R_{b}|^{2}=0.139\pm0.002$ and $|R_{c}|^{2}=0.229\pm0.002$,
leading to extraction efficiency $\eta_{b,\mathrm{ex}}=\frac{\kappa_{b,\mathrm{ex}}}{\kappa_{b}}=0.313\pm0.002$
and $\eta_{c,\mathrm{ex}}=\frac{\kappa_{c,\mathrm{ex}}}{\kappa_{c}}=0.261\pm0.001$
respectively. Thus the on-chip efficiency can be estimated $\eta=\frac{\kappa_{b,\mathrm{ex}}}{\kappa_{b}}\frac{\kappa_{c,\mathrm{ex}}}{\kappa_{c}}\eta_{\mathrm{i}}=(2.11\pm0.03)\%$.
We also followed the calibration procedure in Ref. {[}7{]}. The complete
conversion matrix is measured as shown in Fig.$\:$4 in the main text,
including optical reflection $S_{\mathrm{oo}}$, microwave reflection
$S_{\mathrm{ee}}$, microwave-to-optical conversion $S_{\mathrm{oe}}$,
and optical-to-microwave conversion $S_{\mathrm{eo}}$. The on-resonance
conversion efficiency is normalized by the off-resonance reflection
amplitude, thus the gain and loss of the measurement circuit are excluded.
Assuming that the conversion efficiency is the same for both directions,
we estimate $(2.05\pm0.04)\%$ on-chip efficiency, which agrees well
with the estimation based on the cooperativity and extraction efficiency.

Based on the cooperativity $C$ and the resonance linewidth, the enhanced
coupling strength can be estimated $G_{\mathrm{eo}}=2\pi\times(1.76\pm0.05)\:\mathrm{MHz}$.
The pump photon number inside the pump optical mode is $(3.2\pm0.1)\times10^{7}$,
with the uncertainty determined by the insertion loss (supplementary
material of Ref. {[}34{]}). Therefore the effective vacuum coupling
rate estimated from experimental results is 
\begin{equation}
g_{\mathrm{eo,exp}}^{\mathrm{eff}}=2\pi\times(310\pm10)\,\mathrm{Hz}.
\end{equation}
Using finite-element-method simulation with the actual device geometry
estimated from Fig.$\,$\ref{FigS5.1-1}, the field distributions
($u_{a}$, $u_{b}$, $u_{c}$) of the optical pump and signal modes,
and the microwave mode can be obtained (Fig.$\,$1D, E, \& F). Plugging
the field distributions into Eq.$\,$(\ref{eq:S1.2-1-1}) and assuming
$r_{113}=1\,\mathrm{pm/V}$ {[}21{]}, the vacuum coupling rate for
our device can be calculated as $g_{\mathrm{eo}}=2\pi\times(520\pm50)\,\mathrm{Hz}$,
with the uncertainty determined by the parameter difference between
the design and fabricated device estimated from Fig$\,$\ref{FigS5.1-1}.
Together with the optical mode mixing strength obtained from Eq.$\,$(\ref{eq:S3.2-1}),
the calculated effective vacuum coupling rate is 
\begin{equation}
g_{\mathrm{eo}}^{\mathrm{eff}}=g_{\mathrm{eo}}\cos(2\theta)=2\pi\times(330\pm30)\,\mathrm{Hz},
\end{equation}
which agrees with our experimental estimation. 

\section{Added noise during conversion}

In addition to the conversion efficiency $\eta$, the added noise
during the conversion process is also an important figure of merit.
There are mainly two sources of noise, thermal excitation of the microwave
cavity and photons generated by the parametric interaction $\hbar G(b^{\dagger}c^{\dagger}+bc)$.

By including the counter-rotating term $b^{\dagger}c^{\dagger}+bc$
, we have the 
\begin{align}
 & -i\omega b\left(\omega\right)=-\left(i\delta_{b}+\frac{\kappa_{b}}{2}\right)b\left(\omega\right)-iGc\left(\omega\right)-iGc^{\dagger}\left(-\omega\right)+\sqrt{\kappa_{b,\mathrm{ex}}}b_{\mathrm{in}}\left(\omega\right)\delta\left(\omega-\delta_{b,in}\right)+\sqrt{\kappa_{b,\mathrm{in}}}b_{n}\left(\omega\right),\label{eq:S21}\\
 & -i\omega b^{\dagger}\left(-\omega\right)=-\left(-i\delta_{b}+\frac{\kappa_{b}}{2}\right)b^{\dagger}\left(-\omega\right)+iGc\left(\omega\right)+iGc^{\dagger}\left(-\omega\right)+\sqrt{\kappa_{b,\mathrm{ex}}}b_{\mathrm{in}}^{\dagger}\left(\omega\right)\delta\left(-\omega-\delta_{b,in}\right)+\sqrt{\kappa_{b,\mathrm{in}}}b_{n}^{\dagger}\left(-\omega\right),\label{eq:S22}\\
 & -i\omega c\left(\omega\right)=-\left(i\omega_{c}+\frac{\kappa_{c}}{2}\right)c\left(\omega\right)-iGb\left(\omega\right)-iGb^{\dagger}\left(-\omega\right)+\sqrt{\kappa_{c,\mathrm{ex}}}c_{\mathrm{in}}\left(\omega\right)\delta\left(\omega-\omega_{c,in}\right)+\sqrt{\kappa_{c,\mathrm{in}}}c_{n}\left(\omega\right),\label{eq:S23}\\
 & -i\omega c^{\dagger}\left(-\omega\right)=-\left(-i\omega_{c}+\frac{\kappa_{c}}{2}\right)c^{\dagger}\left(-\omega\right)+iGb\left(\omega\right)+iGb^{\dagger}\left(-\omega\right)+\sqrt{\kappa_{c,\mathrm{ex}}}c_{\mathrm{in}}^{\dagger}\left(-\omega\right)\delta\left(-\omega-\omega_{c,in}\right)+\sqrt{\kappa_{c,\mathrm{in}}}c_{n}^{\dagger}\left(-\omega\right),\label{eq:S24}
\end{align}
Here, $b_{n}\left(\omega\right)$ and $c_{n}\left(\omega\right)$
are the noise from the intrinsic loss channels of the optical and
microwave cavities respectively. The noise correlators can be expressed
as following
\begin{align}
\left\langle b_{n}^{\dagger}\left(\omega\right)b_{n}\left(\Omega\right)\right\rangle  & =0\\
\left\langle b_{n}\left(\Omega\right)b_{n}^{\dagger}\left(\omega\right)\right\rangle  & =\delta\left(\omega-\Omega\right)\\
\left\langle c_{n}^{\dagger}\left(\omega\right)c_{n}\left(\Omega\right)\right\rangle  & =n_{c}\delta\left(\omega-\Omega\right)\\
\left\langle c_{n}\left(\Omega\right)c_{n}^{\dagger}\left(\omega\right)\right\rangle  & =\left(n_{c}+1\right)\delta\left(\omega-\Omega\right)
\end{align}

By solving Eq. (\ref{eq:S21}) - (\ref{eq:S24}) and the input-output
relations Eq. (\ref{eq:S1.9}) \& Eq. (\ref{eq:S1.10}), we can obtain
the output signal at phase matching condtion $\delta_{b}=\omega_{c}$
\begin{align}
\left\langle b_{\mathrm{out}}^{\dagger}\left(\omega\right)b_{\mathrm{out}}\left(\omega\right)\right\rangle  & =\eta_{c\rightarrow b}\left\langle c_{\mathrm{in}}^{\dagger}\left(\omega\right)c_{\mathrm{in}}\left(\omega\right)\right\rangle +N_{b,add}+N_{b,para},\\
\left\langle c_{\mathrm{out}}^{\dagger}\left(\omega\right)c_{\mathrm{out}}\left(\omega\right)\right\rangle  & =\eta_{b\rightarrow c}\left\langle b_{\mathrm{in}}^{\dagger}\left(\omega\right)b_{\mathrm{in}}\left(\omega\right)\right\rangle +N_{c,add}+N_{c,para},
\end{align}
with the conversion efficiency
\begin{align}
\eta_{c\rightarrow b} & =\frac{\kappa_{b,\mathrm{ex}}\kappa_{c,\mathrm{ex}}}{\kappa_{b}\kappa_{c}}\frac{4C}{\left(1+C\right)^{2}}\nonumber \\
 & +\frac{\kappa_{b,\mathrm{ex}}\kappa_{c,\mathrm{ex}}}{\kappa_{b}\kappa_{c}}\frac{C}{(C+1)^{4}}\left[\left(C^{2}+2C\right)\frac{\kappa_{b}^{2}}{4\omega_{c}^{2}}+\left(1+4C+2C^{2}\right)\frac{\kappa_{c}^{2}}{4\omega_{c}^{2}}+2C\frac{\kappa_{c}\kappa_{b}}{4\omega_{c}^{2}}\right]+O(1/\omega_{c}^{3}),\\
\eta_{b\rightarrow c} & =\frac{\kappa_{b,\mathrm{ex}}\kappa_{c,\mathrm{ex}}}{\kappa_{b}\kappa_{c}}\frac{4C}{\left(1+C\right)^{2}}+\nonumber \\
 & +\frac{\kappa_{b,\mathrm{ex}}\kappa_{c,\mathrm{ex}}}{\kappa_{b}\kappa_{c}}\frac{C}{(C+1)^{4}}\left[\left(C^{2}+2C\right)\frac{\kappa_{c}^{2}}{4\omega_{c}^{2}}+\left(1+4C+2C^{2}\right)\frac{\kappa_{b}^{2}}{4\omega_{c}^{2}}+2C\frac{\kappa_{c}\kappa_{b}}{4\omega_{c}^{2}}\right]+O(1/\omega_{c}^{3}),
\end{align}
The added noise due to the thermal environment is 
\begin{align}
N_{b,add} & =\frac{4C}{\left(1+C\right)^{2}}\frac{\kappa_{b,\mathrm{ex}}\kappa_{c,\mathrm{in}}}{\kappa_{b}\kappa_{c}}n_{c},\\
N_{c,add} & =\frac{4}{\left(1+C\right)^{2}}\frac{\kappa_{c,\mathrm{in}}\kappa_{c,\mathrm{ex}}}{\kappa_{c}^{2}}n_{c}.
\end{align}
And the noise due the counter-rotating term is
\begin{align}
N_{b,para} & =\frac{C}{(C+1)^{2}}\frac{\kappa_{b,\mathrm{ex}}}{\kappa_{b}}\frac{C\kappa_{b}^{2}+\kappa_{c}^{2}}{4\omega_{c}^{2}},\\
N_{c,para} & =\frac{C}{(C+1)^{2}}\frac{\kappa_{c,\mathrm{ex}}}{\kappa_{c}}\frac{\kappa_{b}^{2}+C\kappa_{c}^{2}}{4\omega_{c}^{2}}
\end{align}

Compare with Eq.$\,$(\ref{eq:S1.13}), the additional term in the
conversion efficiency is the counter-rotating term induced amplification.
For our experiment parameters with $\kappa_{b}/2\pi=294\,\mathrm{MHz}$,
$\kappa_{c}/2\pi=0.55\,\mathrm{MHz}$, and $\omega_{c}/2\pi=8.31\,\mathrm{GHz}$,
we estimated the modification of the conversion efficiency $\eta\rightarrow\eta\left(1+\mathcal{A}\right)$
due to the parametric amplification factor $\mathcal{A}\sim\frac{\left(C^{2}+2C\right)}{4\left(1+C\right)^{2}}\frac{\kappa_{b}^{2}}{4\omega_{c}^{2}}=1.0\times10^{-5}$
for the achieved $C\sim0.075$. This indicates that the RWA is valid
for our experiments. 

For the added noise, we have the thermal excitation $n_{c}\sim4.5$
for $T\sim2\,\mathrm{K}$, thus $N_{b,add}\sim0.3$ and $N_{c,add}\sim3$,
which can be further reduced by working at lower temperature. In addition,
the noise due to parametric amplification can be estimated, $N_{b,para}\sim5\times10^{-7}$
and $N_{c,para}\sim5\times10^{-6}$, which are negligible in our current
experiments.
\end{document}